\documentclass[doublecol]{epl2} 
\usepackage[utf8x]{inputenc}
\usepackage{graphicx}
\usepackage{stackrel}
\usepackage{epsfig}
\usepackage{amsmath}
\usepackage{amsfonts}
\usepackage{amssymb}

\title{Normal state magnetotransport properties of $\beta$-FeSe superconductors}
\shorttitle{Normal state magnetotransport properties of $\beta$-FeSe} %Insert here a short version of the title if it exceeds 70 characters

\author{J. D. Querales-Flores\inst{1,2} \and M. L. Amig\'o \inst{1,2} \and G. Nieva \inst{1,2} \and C. I. Ventura\inst{1,3} }
\shortauthor{J. D. Querales \etal}

\institute{                    
  \inst{1} Centro At\'omico Bariloche-CNEA and CONICET, Av. Bustillo 9500, 8400 Bariloche, Argentina \\
  \inst{2} Instituto Balseiro, Univ. Nac. de Cuyo and CNEA, 8400 Bariloche, Argentina\\
  \inst{3} Sede Andina, Univ. Nac. de R\'io Negro, 8400 Bariloche, Argentina
  
}

\abstract{
We present $\beta$-FeSe magnetotransport data, and describe them theoretically. Using a simplified microscopic model with two correlated effective orbitals, we determined the normal state electrical conductivity and Hall coefficient, using Kubo formalism. With model parameters relevant for Fe-chalcogenides, we describe the observed effect of the structural transition on the ab-plane electrical resistivity, as well as on the magnetoresistance. Temperature-dependent Hall coefficient data were measured at 16 Tesla, and their theoretical description improves upon inclusion of moderate electron correlations. We confirm the effect of the structural transition on the electronic structure, finding deformation-induced band splittings comparable to those reported in angle-resolved photoemission.}

\begin{document}

\maketitle

\section{Introduction}

Since the discovery of superconductivity in LaFeAsO$_{1-x}$F$_{x}$,\cite{kamihara2008} several types of iron-based superconductors have been reported. The so called ``11" family of FeSe superconductors attracted much attention due to their simpler crystal structure, and particular electronic and physical properties. 
Since the first report of superconductivity with critical temperature $T_{c} = 8$K for PbO-type $\alpha$-FeSe$_{0.88}$ by Hsu et al.\cite{hsu2008},  a $T_{c}$ of 37 K 
at a pressure of 8.9 GPa was already  reached.\cite{medvedev} FeSe compounds have a similar band structure to ferropnictides.\cite{subedi2008,review2015}
FeSe$_{1-x}$ with Se deficiency was reported to exhibit anomalies related to spin density waves (SDW) and magnetic ordering at temperatures near 100 K.\cite{lee2008} On the other hand, Ref.\cite{mcqueen2009} reported that FeSe exhibited superconductivity within a narrow range of stoichiometries, Fe$_{1.01\pm0.02}$Se, without magnetic ordering.

Pure $\beta$-FeSe undergoes a structural transition from a low-temperature orthorhombic to a tetragonal phase at $T_{s}\sim 90$K, not accompanied by a SDW, 
and the compound exhibits superconductivity below $T_{c} = 8.87$K.
 %In this system the orthorrombic distortion and superconductivity do not compete\cite{bohmer2013}. 
Angle-resolved photoemission spectroscopy (ARPES) experiments in  $\beta$-FeSe revealed a significant change in the electronic structure when going through the structural transition.\cite{shimojima2014} Recently it was claimed \cite{watson2015} that the observed changes in electronic structure could not be explained by the small lattice distortion, issue which we will address in our present work.  

Recently, Amig\'o et al.\cite{amigo2014} reported that multiband effects are needed to describe the magnetotransport properties of  $\beta$-FeSe (Fe$_{0.96}$Se) single crystals. Concretely, in the normal state below 90 K, a strongly anisotropic positive magnetoresistance, that becomes negligible above that temperature, was found. 
This magnetoresistance and the upper critical field could be understood with a phenomenological uncorrelated two-band model. Also a recent ultra-high magnetic field study\cite{watson2-2015} reported that magnetotransport in FeSe results from a small multiband Fermi surface (FS) with different carrier mobilities.

In this work, to study normal state magnetotransport properties of $\beta$-FeSe superconductors, we propose to employ a minimal microscopic model, which includes two effective bands describing the low-energy electronic structure, as well as intra- and inter-orbital Coulomb interactions. Previously\cite{condmat2015}  we treated the model 
using perturbative techniques to determine the electron Green's functions and the temperature-dependent  spectral density function. 
The kinetic energy part of the Hamiltonian is represented by the effective two-orbital model proposed by Raghu et al. in Ref.\cite{raghu}, consisting of 
a two-dimensional lattice for the Fe atoms, with two degenerate orbitals per site. Tight-binding parameters were fitted 
to obtain an effective band structure describing the Fermi surface topology of ferropnictides.\cite{raghu,yao2009} The two-orbital model 
was shown to be suitable to describe the  extended s-wave pairing and other superconducting properties of these systems.\cite{Graser2009,yao2009,Ran2009,Hu2009,zhou2011,dagotto2011,jhu2012,gliu2014} 

%In the following, we present the microscopic model here used to describe FeSe compounds, and the theoretical approach adopted to calculate magnetotransport properties 
%and analize the effect of the structural transition  on the  electronic properties of FeSe. 
%Next, we present experimental magnetotransport results and their theoretical description. We compare the parameters used in the phenomenological approach 
%of Ref.\cite{amigo2014}, with those of our correlated two-orbital model model. We also compare our calculated deformation-induced band splittings with recent ARPES data, before concluding. 

\section{Calculation of magnetotransport properties of FeSe compounds}\label{section2}

 \subsection{Microscopic two-orbital minimal model for FeSe} 
  To describe analytically the normal state magnetotransport properties of FeSe superconductors, we will consider the following minimal model preserving the essential low-energy physics: 
\begin{eqnarray}
 \mathcal{H} & = & \mathcal{H}_{0} + V_{int}
 \label{Hamiltonian}
 \end{eqnarray}

\noindent The kinetic energy part of the Hamiltonian in Eq. \ref{Hamiltonian} is given by the uncorrelated two-orbital model  by Raghu et al.\cite{raghu} mentioned in the Introduction:
\begin{eqnarray}
 {\mathcal{H}}_{0} &  = &\sum_{k,\sigma}{\left[{E}_{c}(k) {c}^{\dagger}_{k\sigma} {c}_{k\sigma} + {E}_{d}(k) {d}^{\dagger}_{k\sigma}{d}_{k\sigma}\right]}
  \label{Hamiltonian0}
\end{eqnarray}
where ${c}^{\dagger}_{k\sigma}$ creates an electron with crystal momentum $\vec{k}$ and spin $\sigma$ in the effective band with energy $E_{ c } (\vec{k})$, 
likewise for  ${d}^{\dagger}_{k\sigma}$ and $E_{ d } (\vec{k})$. The effective band energies are:
\begin{equation}
  E_{  \overset { d }{ c } }(\vec{k}) = \epsilon_{+}(\vec{k}) \pm \sqrt{ \epsilon_{-}^{2}(\vec{k}) + \epsilon_{xy}^{2}(\vec{k})} -\mu
  %- \mu  
  % \nonumber
  \label{effectivebands}
 \end{equation}

\noindent $\mu$ denotes the chemical potential at temperature $T$, and:
%notice that we refer the energies to $\mu$, since we will adopt the grand canonical ensemble 
%in our Green's function treatment of the system, like  in  Ref.~\cite{raghu}.  
%$\sigma$ denotes the spin degrees of freedom, and 

 \begin{eqnarray}
  \epsilon_{\pm}(\vec{k})&=& \frac{\epsilon_{x}(\vec{k}) \pm \epsilon_{y}(\vec{k}) }{2}; \,\,\,\, \epsilon_{xy}(\vec{k}) =  -4t_{4}\sin(k_{x})\sin(k_{y}) \nonumber\\
  \epsilon_{x}(\vec{k}) & = & -2t_{1}\cos(k_{x}) - 2t_{2}\cos(k_{y})- 4t_{3}\cos(k_{x})\cos(k_{y})  \nonumber \\
  \epsilon_{y}(\vec{k}) & = & -2t_{2}\cos(k_{x}) - 2t_{1}\cos(k_{y}) -4t_{3}\cos(k_{x})\cos(k_{y}) \nonumber
\end{eqnarray}

\noindent  The tight-binding parameters $t_{i}, i=1-4, $ denote the hopping amplitudes between sites of the two-dimensional lattice of Fe atoms, 
derived in Ref. \cite{raghu}  as: $ t_{1} = -1 $ eV, $t_{2}  = 1.3 $ eV , $t_{3}$  = $t_{4}  = -0.85 $ eV.

The electron correlations are represented by $V_{int}$ in Eq. \ref{Hamiltonian}. The effect of local intra- and inter-orbital correlations 
in ferropnictides was previously studied.\cite{dagotto2011,scalapino2012,condmat2015} It was found that the inter-orbital correlation was less relevant 
than the intra-orbital one. Therefore, in our minimal model for FeSe we consider only the local intra-orbital Coulomb repulsion $U$: 
\begin{equation}
 {V}_{int}  =  \sum_{i} U \left( {n}_{i\uparrow}{n}_{i\downarrow} +{N}_{i\uparrow}{N}_{i\downarrow}   \right) 
%\raisetag{1pt}
 \label{Vint}
 \end{equation}
where: $n_{i\sigma} = {c}^{\dagger}_{i\sigma}{c}_{i\sigma}$ and $N_{i\sigma} = {d}^{\dagger}_{i\sigma}{d}_{i\sigma}$, and $i$ denotes the Fe-lattice sites.  Since 
correlations in FeSe compounds are intermediate,\cite{aichhorn2010, craco1-2014, craco2-2014, maletz2014, condmat2015,vollhardt2015} and mainly motivated by 
the fact that it had been possible to describe previous magnetotransport results in terms of a phenomenological model with two uncorrelated carrier bands,\cite{amigo2014} 
 here we decided to use Hartree-Fock approximation (HF) for the correlations. A recent study of the effect of correlations 
 in FeSe\cite{vollhardt2015}, which found no relevant qualitative differences employing density functional theory (DFT) calculations  and DFT+DMFT (DFT with 
 dynamical mean field theory) for  the FS and the low energy spectral properties, provides further justification for the level of approximation we used.  
We determined the HF renormalized band structure, and self-consistently calculated $\mu(T)$ for total electron filling $n$ of the two renormalized effective bands (see Ref.\cite{condmat2015} for details).

%\end{widetext}

\subsection{Calculation of the electrical conductivity tensor and Hall coefficient}
\label{calculations}

To describe magnetotransport in FeSe compounds, we evaluated the electrical conductivity tensor $\sigma_{\alpha\beta}$, defined by: 
\begin{equation}
\langle j_{\alpha}(t)\rangle = \sigma_{\alpha\beta}E_{\beta}(t)
\end{equation}

\noindent where $\langle j_{\alpha}(t)\rangle$ is the average current at temperature $T$ and time $t$ flowing in the $\alpha$-direction, in response to an electric field, $E_{\beta}(t)$, applied in the $\beta$-direction. 

Assuming the presence of a magnetic field $\vec{H} = H_z \hat{z} $ perpendicular to the ab-plane of FeSe, and the electric current flowing in the $x$-direction ($j_x$) as a  
result of an electric field along $\hat{x}$ plus the Hall electric field along $\hat{y}$:
\begin{equation}
 \langle j_{x} \rangle  = \sigma_{xx}(\omega) E_{x}(t) + \sigma_{xy}(\omega) E_{y}(t)  
 \end{equation}

\noindent where $\sigma_{xx}(\omega)$ and $\sigma_{xy}(\omega)$, are respectively the longitudinal and transversal components of the electrical conductivity tensor. 
To compare our analytical results with experiments, we  determined the ab-plane dc-resistivity ($\rho_{xx}$) and the Hall resistivity ($\rho_{xy}$) as the static (zero-frequency,  i.e $\omega \to 0$) limit of:
\begin{equation}
\rho_{xx} = \frac{\sigma_{xx}(\omega)}{\sigma_{xx}^{2}(\omega) + \sigma_{xy}^{2}(\omega)}; \,\,\,\,\,\,\,\, \rho_{xy} = \frac{\sigma_{xy}(\omega)}{\sigma_{xx}^{2}(\omega) + \sigma_{xy}^{2}(\omega)} 
\end{equation}

 In the Kubo formulation for transport,\cite{kubo,stinchcombe} $ \sigma_{\alpha\beta}$ are given by appropriate generalised susceptibilities $\chi_{AB}(\omega)$, 
measuring the linear response of observable $A$ of a system to an applied external field coupling to its observable $B$. The susceptibilities, in turn, can be 
calculated using retarded Green's functions, $\ll A; B \gg (\omega)$.\cite{zubarev, stinchcombe} Here:
 \begin{eqnarray}
  \sigma_{xx}(\omega) & = \chi_{j_{x},eX}(\omega) & =  \ll j_{x}; eX \gg(\omega)  \\ 
  \sigma_{xy}(\omega) & = \chi_{j_{x},eY}(\omega) & =  \ll j_{x}; eY \gg(\omega)
 \end{eqnarray}
\noindent where $X$ and $Y$ are the respective components of the system's position operator.
%Moreover, the generalized susceptibilities $\chi_{J_{x},eX}$ and $\chi_{J_{x},eY}$ are given by the following retarded Green's functions\cite{zubarev, stinchcombe} :
%\begin{eqnarray}\label{greens}
%&\chi_{j_{x},eX} = \ll j_{x}; eX \gg(\omega + i\delta);\nonumber \\
%& \chi_{j_{x},eY} = \ll j_{x}; eY \gg(\omega + i\delta),
%\end{eqnarray}
The electron Green's functions include a sum of respective contributions from the $c$ and $d$ effective bands, which 
%\begin{align}\label{greens1}
%&\chi_{j_{x},eX} = \ll j_{x}, eX \gg_{c}(\omega + i\delta) + \ll j_{x}, eX \gg_{d}(\omega + i\delta) \nonumber \\
%&\chi_{j_{x},eY} = \ll j_{x}, eY \gg_{c}(\omega+i\delta) + \ll j_{x}, eY \gg_{d}(\omega+i\delta)
%\end{align}
 can each be calculated from the following exact set of equations of motion (EOM)\cite{zubarev}:
\begin{align}\label{greens}
& \omega  \ll j_{x}, eX \gg^{c,d}  = \frac{1}{2\pi}\langle [j^{c,d}_{x},eX] \rangle + \ll [j^{c,d}_{x}, \mathcal{H} ]; eX \gg \nonumber \\ 
& \omega  \ll j_{x}, eY \gg^{c,d}   = \frac{1}{2\pi}\langle [j^{c,d}_{x},eY] \rangle + \ll [j^{c,d}_{x}, \mathcal{H} ]; eY \gg
\end{align}
\noindent where the current operator\cite{mahan} is defined as: $j^{c}_{x} = \frac{e}{m^{*}_{c}}\sum_{\vec{k},\sigma}{k_{x}c^{\dagger}_{\vec{k},\sigma}c_{\vec{k},\sigma}}$ and $j^{d}_{x} = \frac{e}{m^{*}_{d}}\sum_{\vec{k},\sigma}{k_{x}d^{\dagger}_{\vec{k}, \sigma}d_{\vec{k},\sigma}}$, being $ m^{*}_{i}$, i=c,d, the effective masses of the carriers 
in each band.  New higher order Green's functions appear coupled in Eqs. \ref{greens}. In order to close the system of coupled equations of motion 
we used HF approximation to decouple them, 
and determined $ \ll j_{x}, eX \gg $ and $  \ll j_{x}, eY \gg $ in first order of perturbations on the electron correlation $U$. 
The final expressions obtained for the ab-plane electrical conductivity components, in presence of $\vec{H} = H_z \hat{z}$, read:
\begin{eqnarray}\label{sigmaxx}
\sigma_{xx}(\omega) = \frac{e^{2}}{\Omega} \sum_{\vec{k},\sigma}\left\lbrace  \frac{\langle c^{\dagger}_{\vec{k}\sigma}c_{\vec{k}\sigma}\rangle}{\hbar(\omega - \omega_{c})-n_{c}E_{c}(\vec{k})-2Un_{c}^{2}} \right. \nonumber \\
\left. +  \frac{\langle d^{\dagger}_{\vec{k}\sigma}d_{\vec{k}\sigma}\rangle}{\hbar(\omega - \omega_{d})-n_{d}E_{d}(\vec{k})-2Un_{d}^{2}}   \right\rbrace
\end{eqnarray}

\begin{align} \label{sigmaxy}
& \sigma_{xy}  (\omega)  =  \frac{ne}{H_{z}} +  \frac{e^{2}}{\Omega}\sum_{\vec{k},\sigma} \phi(\vec{k})  \left\lbrace \frac{1}{\hbar\omega  - \tilde{E_{c}}(\vec{k}) + \hbar(\omega+\omega_{c})  }\right.\nonumber \\
& \left. -  \frac{1}{-\hbar\omega - \tilde{E_{c}}(\vec{k}) +  \hbar(\omega-\omega_{c}) }  +  \frac{1}{\hbar\omega - \tilde{E_{d}}(\vec{k}) + \hbar(\omega+\omega_{d})  }\right. \nonumber \\
&\left.  -  \frac{1}{-\hbar\omega - \tilde{E_{d}}(\vec{k}) +  \hbar(\omega-\omega_{d})  }  \right\rbrace 
\end{align}

\noindent where: $\Omega$ is the unit cell volume, $ \tilde{E_{i}}(\vec{k}) =  E_{i}(\vec{k}) + 2 U n^{2}_{i} $ for $i= c,d$. Above: 
$\phi(\vec{k}) \equiv \left(\frac{\langle c^{\dagger}_{\vec{k}\sigma}c_{\vec{k}\sigma}\rangle - \langle d^{\dagger}_{\vec{k}\sigma}d_{\vec{k}\sigma}\rangle}{E_{d}(\vec{k})-E_{c}(\vec{k})}\right)$,  being $\omega_{i} \equiv \frac{eH_{z}}{c}\left(\frac{1}{m^{*}_{i}}\right)$ ($i=c,d$), i.e. the cyclotron frequency of $c$ and $d$ electrons. 
$m^{*}_{i}$, $i=c,d$ represent the diagonal components of the effective mass tensor, given by:
$
\left(\frac{1}{m^{*}_{i}}\right)_{\mu\nu} = \frac{1}{\hbar^{2}} \frac{\partial^{2}E_{i}(\vec{k})}{\partial{k_{\mu}}\partial{k_{\nu}}}
$. The conductivity due to multiple band maxima or minima is proportional to the sum of the inverse of the individual masses, 
multiplied by the density of carriers in each band, to take into account all contributions to the conductivity.\cite{cardona} To evaluate the conductivities, we used  
the Chadi-Cohen BZ sampling method\cite{chadicohen,cunningham} for square and rectangular lattices, 
to perform  the required BZ summations.  

The following expression for the Hall coefficient ($R_{H}$) was obtained,  using Eq.\ref{sigmaxy}:  
\begin{align}
&R_{H}=\frac{1}{\sigma_{xy}H_{z}}; \,\,\,\,\,\, \sigma_{xy} = \lim_{\substack{\omega \to 0 \\ \delta \to 0^{+}}}{\Re\left[ \sigma_{xy}(\omega + i\delta)\right]} \equiv \left(\frac{1}{\gamma_{c} + \gamma_{d}}\right)  \nonumber \\
%\end{eqnarray}
%\noindent where:
%\begin{eqnarray}
&\gamma_{i}\equiv \left\lbrace \left(\tfrac{en_{i}}{m^{*}_{i}}\right)\tfrac{(\omega + \omega_{i})\left[ (\omega - \omega_{i})^{2}  + \delta^{2} \right] + (\omega - \omega_{i})\left[(\omega + \omega_{i})^{2}  + \delta^{2} \right]}{(\omega + \omega_{i})^{2}(\omega - \omega_{i})^{2} + \delta^{2}(\omega + \omega_{i})^{2} - \delta^{2}(\omega - \omega_{i})^{2} + \delta^{4}}\right\rbrace 
\end{align}

In next section, we will compare our Hall coefficient  results  with those obtained using the classical expression for two types of uncorrelated carriers (with charge e):\cite{smith1978}

\begin{equation}\label{eqsingleton}
R_{H} = \frac{1}{e}\frac{(\mu_{c}^{2}n_{c}+\mu_{d}^{2}n_{d})+ (\mu_{c}\mu_{d} H_{z})^{2}(n_{c}+n_{d})}{(\mu_{c}n_{c}+\mu_{d}n_{d})^{2} + (\mu_{c}\mu_{d} H_{z})^{2}(n_{c}+n_{d})^{2}}
\end{equation}

\noindent where $\mu_{i}$ , $i= c, d$,  denotes the mobility in each electron band.  One has: $\mu_{i} = e \tau_{i} / m^{*}_{i}  = \sigma_{i} / (e n_{i})$,\cite{singleton}
 being $\tau^{-1}_{i}$  and  $ \sigma_{i}$ respectively the scattering rates and dc-conductivities  for the electrons in each band.

%In our approach the temperature dependence of the conductivity enters through the filling for each band, $n_{c}$ and $n_{d}$ and the lattice parameters of the Fe square lattice, $a$ and $b$. 

%%%%%%%%%%%%%%%%%

\section{Results and discussion}\label{results}

We present magnetotransport results for the normal state of  FeSe compounds, and compare them with those  calculated as 
presented in previous section. Using the optimal correlation value $ U = 3 eV$, previously  found to describe best other electronic properties of these compounds,\cite{condmat2015}   
 we analize the dependence on temperature, doping and magnetic field $H_{z} = H$, and compare our results with new experimental data and those of Ref.\cite{amigo2014}, 
 as well as with the results obtained assuming uncorrelated electrons.  
 Notice that the value $U=3$eV represents less than one third, $\sim0.29$, of the total bandwidth for uncorrelated electrons,\cite{raghu}
 thus characterising  FeSe compounds as systems with intermediate electron correlations as discussed in previous section.
 %aca completar con refs. que ejemplifiquen U  en FeSe - mats. con correlaciones intermedias : Craco2014, etc

 \begin{figure}[b!]
  \begin{center}
  \includegraphics[width=8.5cm]{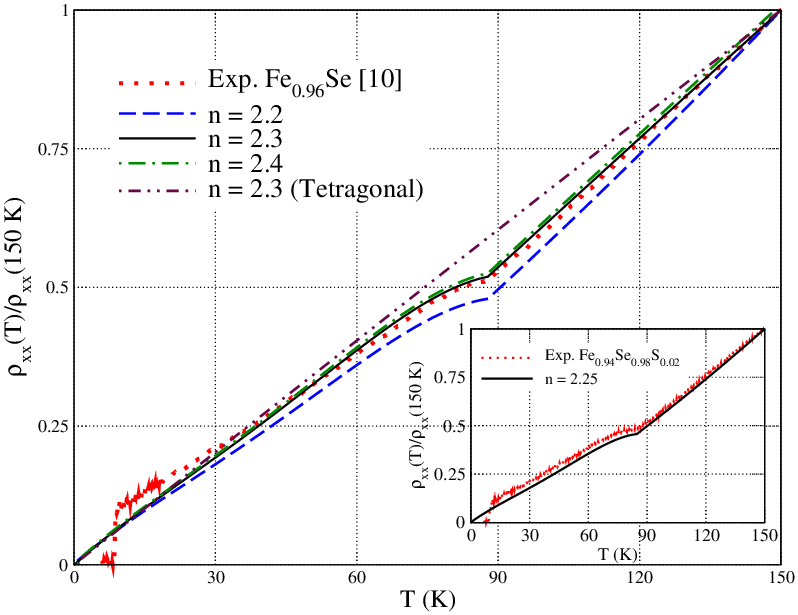}
       \caption[]{$H = 0$ :  ab-plane resistivity as a function temperature.   $\rho_{xx}(T) / \rho_{xx}(150 K)$,  calculated for different doping values (indicated in the figure):
       using the temperature-dependent lattice parameters, $a(T)$ and $b(T)$,   reported for FeSe.\cite{khasanov2010}  Also included is the result obtained 
       assuming a tetragonal lattice, with constant lattice parameter: $ a = b = 3.77 \AA $ ( double dot-dashed line).    
         Experimental curve (dotted line):  Fe$_{0.96}$Se  single crystal,  from Ref.\cite{amigo2014}.       
         Inset: Experimental   $\rho_{xx}(T) / \rho_{xx}(150 K)$ (dotted line) measured for a Fe$_{0.94}$Se$_{0.98}$S$_{0.02}$  single crystal,  and 
         calculated curve(solid line)  for  $n = 2.25$.
           Model parameters used: $U=3$, $t_{1}=-1.0$, $t_{2}=1.3$, $t_{3}=t_{4}=-0.85$. All energies in eV. Chadi-Cohen\cite{chadicohen,cunningham} order 
        for BZ summations: $\nu=9$.}
  \label{figure1}
  \end{center}
  \end{figure}

First, in  Figure \ref{figure1} we study the temperature dependence of the ab-plane dc-resistivity, represented by $\rho_{xx}$(T),  for  Fe$_{0.96}$Se and Fe$_{0.94}$Se$_{0.98}$S$_{0.02}$ single crystals 
in the absence of magnetic field, measured with a standard 4 points dc-technique. The main figure  compares the experimental data  (normalized  at T=150 K)  
with two calculations using our approach:  
one for  a tetragonal crystal  with constant lattice parameters (the normalized resistivity plotted has negligible dependence on doping up to 150 K),
 while  the other, more realistic, takes into account  the T-dependence of the lattice parameters $  a(T), b(T) $  of   FeSe\cite{khasanov2010}  and, in particular, the 
 structural transition,\cite{mcqueen2009-PRL,amigo2014} which  occurs at $T_{s} \sim 90$ K for the Fe$_{0.96}$Se sample, and at 87 K for the  Fe$_{0.94}$Se$_{0.98}$S$_{0.02}$ one.
 As expected,  a clear improvement of the 
 description of  the ab-plane dc-resistivity at $H=0$ is obtained   using the T-dependent lattice parameters of FeSe.\cite{khasanov2010} The best agreement to  the experimental data 
 is obtained  considering  a total electron filling $ n = 2.3$ (main figure),  and $2.25$ (inset), for the  correlated two-orbital model, 
 which corresponds to an Fe-content of x=0.96, and x=0.94, respectively. 
  In accordance with experiment,  the calculated ab-plane resistivity presents 
 a metallic-like behavior in the normal state with a change of slope around the structural transition temperature.
 Hence, we will continue using the temperature-dependent lattice parameters in what follows.
 
    \begin{figure}[t!]
  \begin{center}
  \includegraphics[width=8.5cm]{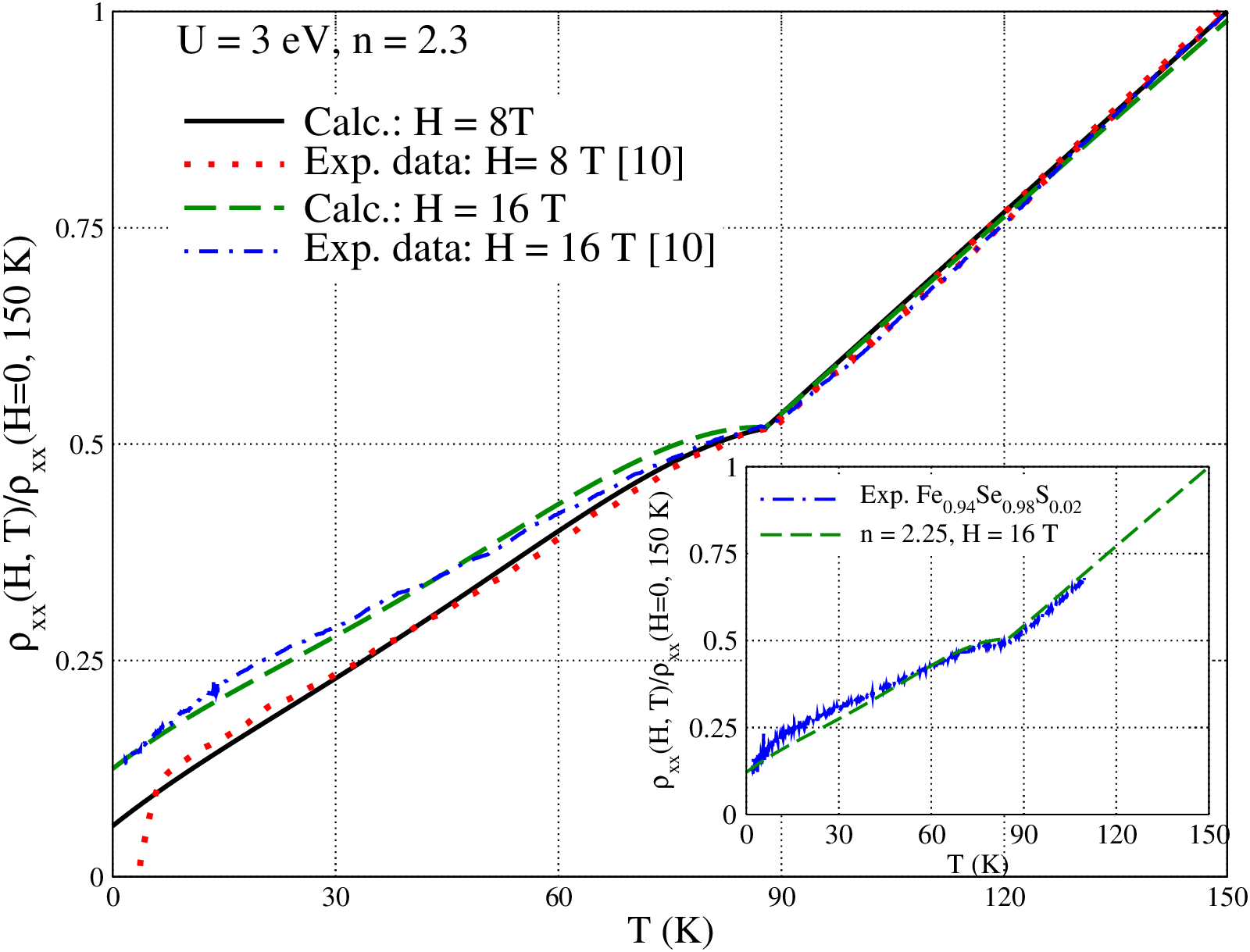}
       \caption[]{Effect of a magnetic field parallel to $c$-axis: temperature dependence of the ab-plane  resistivity (normalized to $\rho(150K, H=0)$) for Fe$_{0.96}$Se 
       ( $n=2.3$ ) in the main figure. Calculated and experimental\cite{amigo2014} results  for H = 8 T, 16 T, as indicated in the plot. Other parameters as in Fig.\ref{figure1}.
       Inset: calculated and experimental\cite{amigo2014} ab-plane  resistivity (normalized to $\rho(150K, H=0)$) of Fe$_{0.94}$Se$_{0.98}$S$_{0.02}$ for H = 16 T.
       }
  \label{figure2}
  \end{center}
  \end{figure}
  
  \begin{figure}[h]
  \begin{center}
   \includegraphics[width=8.5cm]{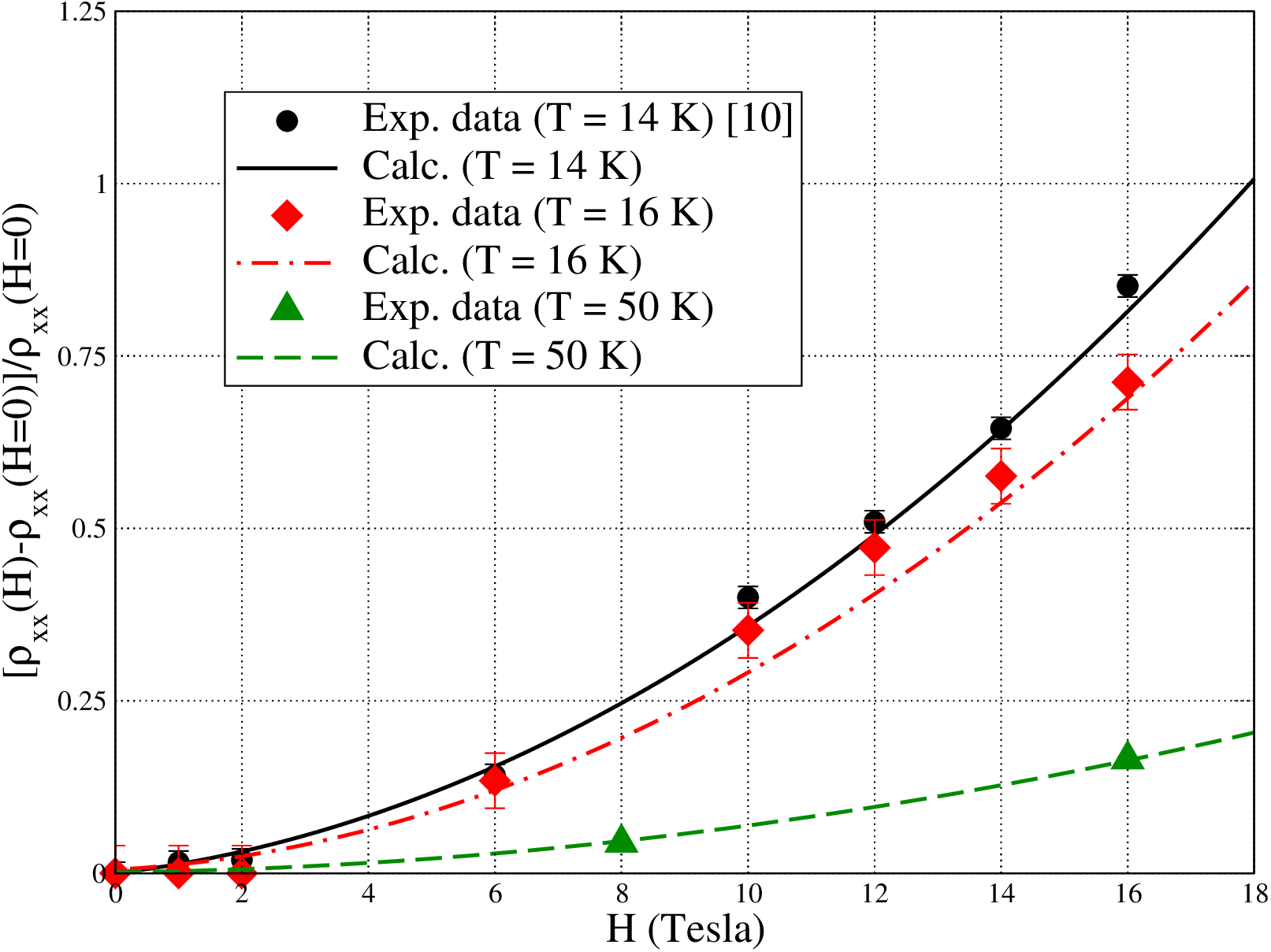}
       \caption[]{ Magnetoresistance as a function of $H$ parallel to the $c$-axis:  calculated (lines) and experimental (symbols) results for temperatures $T=14$, 16,  and $50$ K, 
       as indicated in the plot.
       The experimental data at T=14K are taken from Ref.\cite{amigo2014}.
       Model parameters: $U=3$ eV, $n=2.3$  and others as in Fig.\ref{figure1}. }
  \label{figure3}
  \end{center}
  \end{figure} 
  
In the next three figures we will present magnetotransport results obtained under applied magnetic fields
parallel to the c-axis of the Fe$_{x}$Se  samples:  i.e. perpendicular to the plane formed by the Fe atoms.  
  
In Figure \ref{figure2}, the main figure exhibits the normal state ab-plane resistivity $\rho_{xx}$(T)  calculated and measured at magnetic fields  of 8T  and 16 T, 
having fixed the total band filling 
at $n=2.3$ to describe Fe$_{0.96}$Se. In the inset we show $\rho_{xx}$(T)  at 16 Tesla for the Fe$_{0.94}$Se$_{0.98}$S$_{0.02}$ sample, with the corresponding calculated curve using $n=2.25$. 
Notice  that above $T_{c} = 8.87$ K  for Fe$_{0.96}$Se\cite{amigo2014}, and above $T_{c}= 10.06$ K  for Fe$_{0.94}$Se$_{0.98}$S$_{0.02}$,  we obtain  very good agreement.
 A change of slope of the resistivity at the structural transition temperature is seen, 
 and, in particular, our results describe the positive magnetoresistance observed below $T_{s}$\cite{amigo2014}  and the negligible one above $T_{s}$.

In Figure \ref{figure3} we present  calculated and experimental magnetoresistance results  for Fe$_{0.96}$Se 
 as a function of magnetic field parallel to $c$, at three different temperatures.  
Only the experimental $T = 14 K$ results included have been published before\cite{amigo2014}. 
%   Aqu� alg�n detalle de  su medici�n   exp.   para T= 16  y 50 K  ?
Notice the remarkable agreement at  T=14 K , 16 K , and 50 K  between the experimental magnetoresistance  and the values calculated assuming 
$ U= 3 eV$ and $ n = 2.3$.  
 In particular, our results describe  a quadratic $ \sim H^{2}$ behavior of the magnetoresistance, 
  consistently with the prediction from a phenomelogical two-band model used in Ref.\cite{amigo2014}. 
 In the present work, we also find experimentally and describe theoretically  that the  magnetoresistance concavity (and therefore also  its magnitude)  is monotonically 
 reduced  as temperature is increased  towards $T_{s}\sim90$K, which is consistent with the results in Figure \ref{figure2}, and  in agreement with recent measurements 
 included in an ultra-high magnetic field study of FeSe.\cite{watson2-2015}
 
%  It is interesting to establish a connection between the parameters  of phenomenological two-band fit to magnetotransport data in Ref.\cite{amigo2014} 
%  with those of our correlated two-orbital model.  In Ref.\cite{amigo2014} the magnetoresistance at $T=14$ K, shown in Fig.  \ref{figure3}, was best fitted assuming
%   the following values for the conductivities of each band of uncorrelated electrons: $\sigma_{c}=0.67m\Omega^{-1}cm^{-1}$ and $\sigma_{d}=11.31m\Omega^{-1}cm^{-1}$, 
 %  while our calculations with $U=3 eV$ and $n=2.3$ yield: $\sigma_{c}=0.65m\Omega^{-1}cm^{-1}$ and $\sigma_{d}=11.60m\Omega^{-1}cm^{-1}$.
 % At   $ T = 29.9$ K, the best fit in Ref.\cite{amigo2014} was provided assuming: $\sigma_{c}=0.145m\Omega^{-1}cm^{-1}$ and $\sigma_{d}=6.52m\Omega^{-1}cm^{-1}$, 
 % while in our present approach with $U=3 eV$ and $n=2.3$ we obtain: $\sigma_{c}=0.15m\Omega^{-1}cm^{-1}$ and $\sigma_{d}=6.60m\Omega^{-1}cm^{-1}$. 
 % Therefore, our microscopic model with $U=3 eV$ and $n=2.3$ combined with the analytical approach presented in previous section provides an accurate description 
  %of the magnetotransport data reported for Fe$_{0.96}$Se.\cite{amigo2014}  

At T = 40 K, we find effective masses: $m^{*}_{c}=2.63m_{e}$ and $m^{*}_{d}=3.46m_{e}$, in agreement with DFT+DMFT  calculations by Aichhorn \etal,\cite{aichhorn2010} where a 
significant orbital-dependent mass renormalization in the range of 2 - 5 was predicted, and confirmed by ARPES results at $T=40$K\cite{maletz2014}.

     \begin{figure}[t!]
  \begin{center}
   \includegraphics[width=8.5cm]{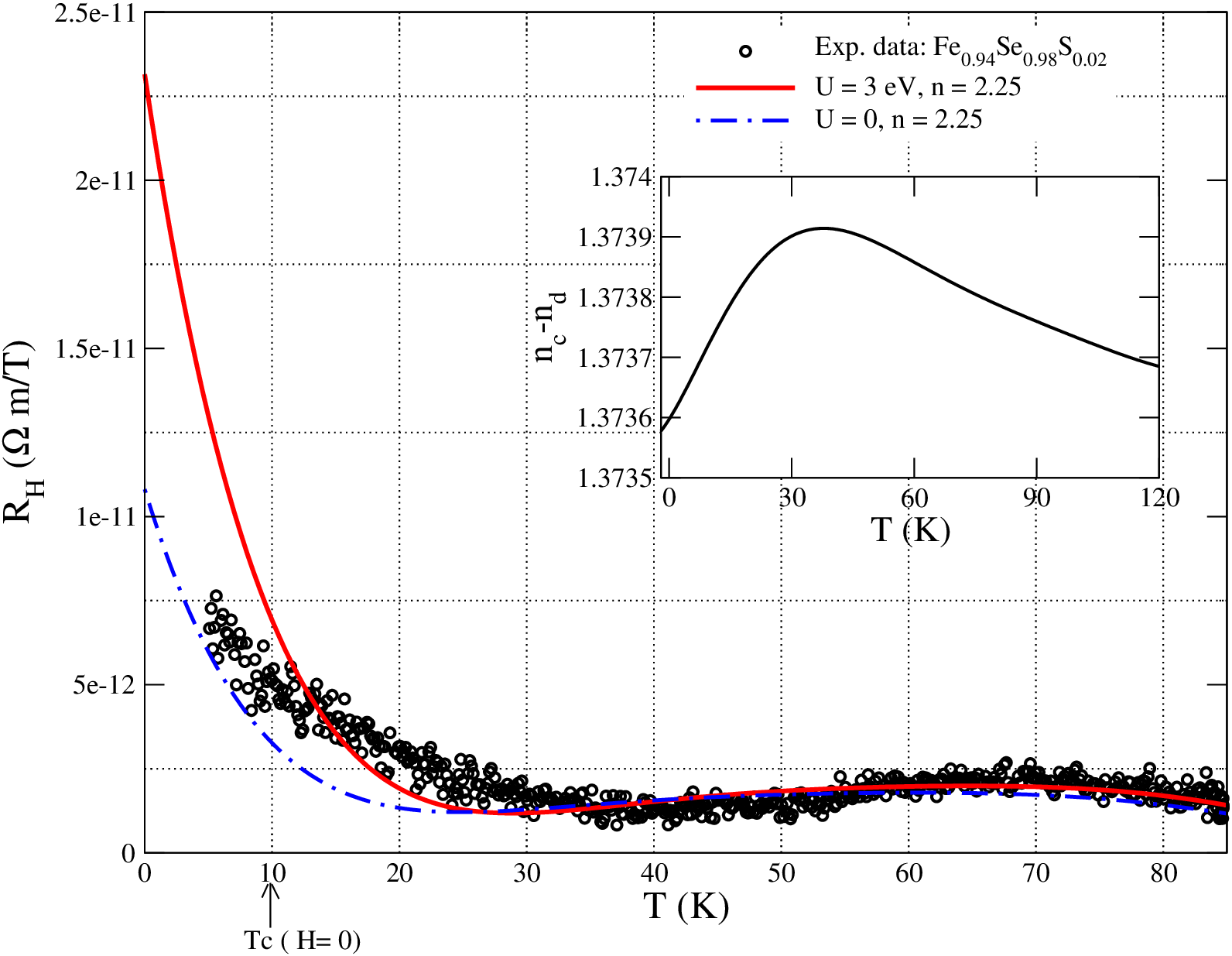}
           \caption[]{Temperature dependence of the Hall coefficient at H = 16 T. Comparison between: our experimental results for Fe$_{0.94}$Se$_{0.98}$S$_{0.02}$ (points), and two theoretical calculations: present 
           analytical approach (solid line) for the correlated two-orbital model( $U= 3 eV, n= 2.25$, other parameters as in Fig.1),  and 
           phenomenological  uncorrelated two-carrier model: Eq.\ref{eqsingleton} (dot-dashed line).  An arrow indicates the critical temperature of the sample at $H=0$.  
           The inset shows  the effect of temperature on the difference ($n_{c} - n_{d}$)  of the partial fillings of the effective bands in our correlated two-orbital model.  }
  \label{Hall}
  \end{center}
  \end{figure}

\begin{figure}[h!]
  \begin{center}
   \includegraphics[width=8.5cm]{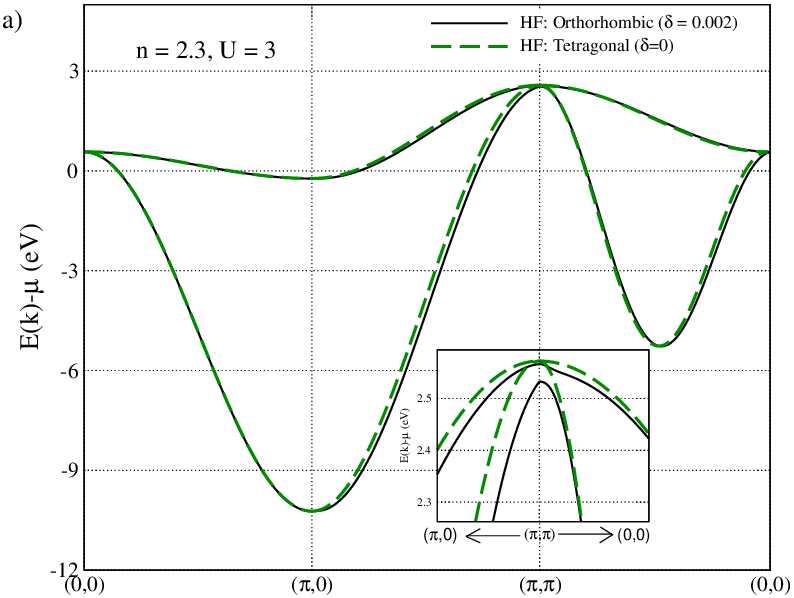}
    \includegraphics[width=8.5cm]{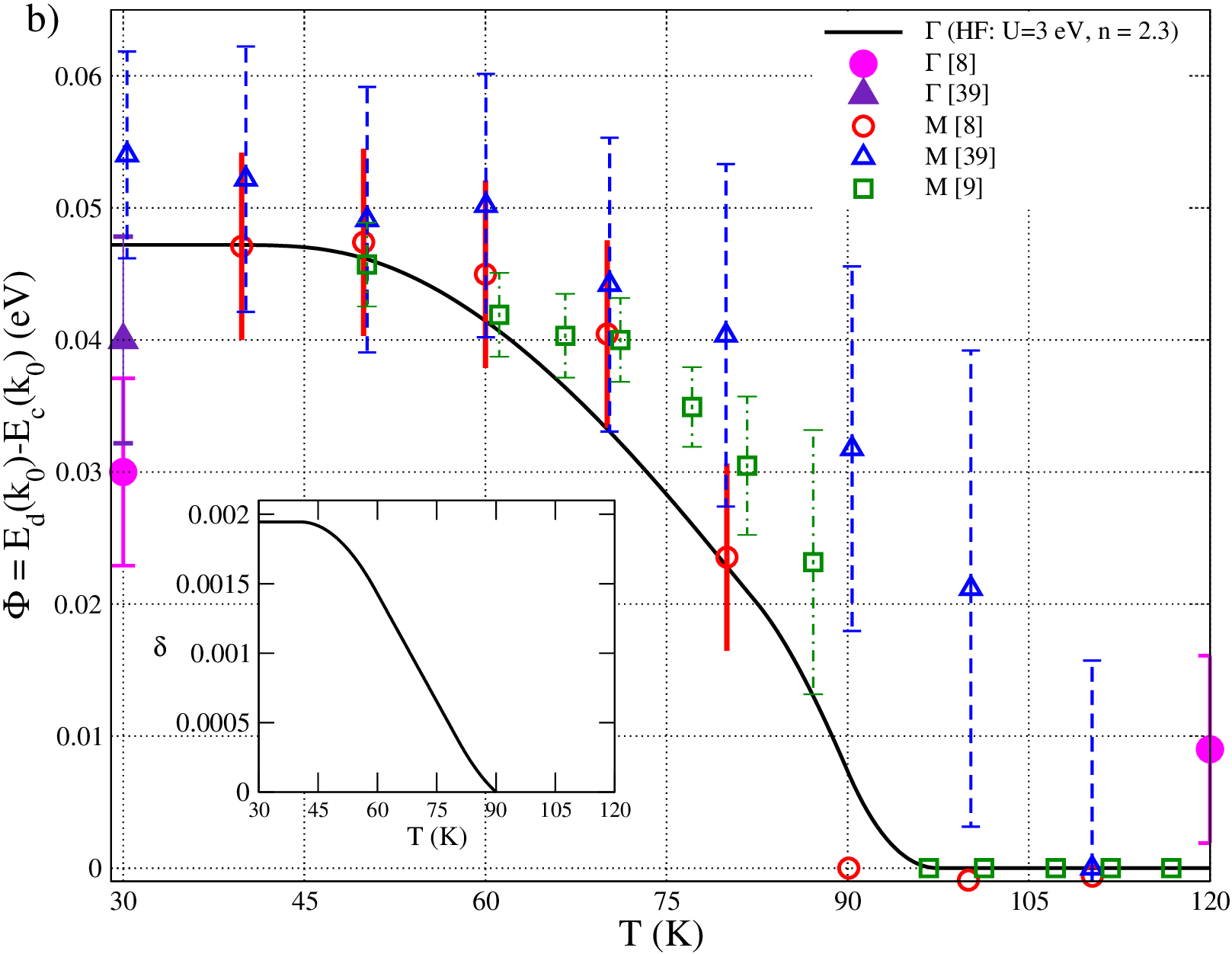}
       \caption[]{H=0, effect of lattice deformation $ \delta$ on the electronic structure.  (a) Band structure of the correlated two orbital model in Hartree-Fock approximation 
       shown in the large (unfolded) BZ,\cite{raghu} i.e. one Fe/cell, at $\delta=0$ (dashed line) and $\delta=0.002$ (solid line). $T = 10 K$, $n_c = 1.87$ and $n_d = 0.43$. 
        Inset: amplification near $\vec{k_0}=(\pi,\pi)$, which corresponds to the zone center $\Gamma$ in the small (folded)  BZ, i.e. two Fe/cell.   (b) T-dependence of the band splitting at two BZ points: denoted as $\Gamma$ and $M$  in the small BZ. Concretely: T-dependence of the calculated band splitting at  $\Gamma$, and for comparison we include respective ARPES data  at $\Gamma$ and $M$. Inset: temperature dependence of the deformation parameter using the lattice parameters of Ref.\cite{khasanov2010}.}
  \label{deformation}
  \end{center}
  \end{figure}

  Next, in Figure \ref{Hall}, we present experimental and theoretical results  obtained for  the Hall coefficient $R_{H}$ in a Fe$_{0.94}$Se$_{0.98}$S$_{0.02}$ single crystal 
  as a function of temperature,  at $H=16$T parallel to the $c$-axis.
   The Hall contribution was measured with a standard dc technique using four contacts 
   along two perpendicular lines, separating the small resistivity contributions by measuring in positive and negative magnetic fields along the  $c$-axis.  
 % The experimental data  were obtained from a Fe$_{0.94}$Se$_{0.98}$S$_{0.02}$ single crystal  (whose compositions were determined by scanning electron microscopy, SEM). 
  We also included in Figure \ref{Hall} the theoretical result obtained with our analytical approach, 
   for the correlated two-orbital model with parameters $U=3 eV$ and filling $n = 2.25 $.
  Notice the good agreement obtained with the experimental  data.
  %( taking into account the experimental error  inherent to the SEM determination of the Fe-composition).  
  We found that in our theoretical approach  $R_{H}$, apart from its dependence on magnetic field, is very sensitive to total electron filling $n$, 
  presenting qualitative sizeable changes depending on the Fe-content. These changes are related to the 
  position of the Fermi level with respect  to the effective model's band structure\cite{raghu,condmat2015} (which can be seen in Fig.\ref{deformation}(a)).
  The theoretical curve  in Figure \ref{Hall}   corresponds to a multi-band situation  in which the Fermi level crosses the two $c$ and $d$ correlated bands, 
  with unequal fillings of those bands.  In particular, the inset depicts the temperature dependence of the  difference ($n_{c} - n_{d}$)  
  between the partial fillings of these bands at total filling $n=2.25$. Notice  that it is maximum at the same temperature,
   $ \sim 38 K$, at which the dependence on temperature of the lattice parameters sets in. This maximum coincides with the inflection point in  $R_{H} (T)$, which we checked 
   that also happens at $ H = 16 T$  if two uncorrelated carrier bands contributed to $R_{H} (T)$  according to Eq.\ref{eqsingleton}.
   The latter case is also shown  in Figure \ref{Hall},  using the carrier mobilities and densities obtained from our approach for: $U = 0$ and $ n = 2.25$. 
    Figure \ref{Hall} evidentiates  that  better agreement  to the experimental data  is obtained  with the correlated two-orbital model, than in the absence of electron correlations.

To end, we discuss  the effect of the lattice deformation related to the structural transition  on the  electronic properties of FeSe superconductors, 
 in the absence of magnetic field. It has been suggested  that the emergence of magnetoresistance in FeSe superconductors below $T_{s}$ 
might be related to changes in the electronic structure.\cite{amigo2014,shimojima2014} 
On the HF renormalized band structure of our effective correlated two-orbital model for FeSe compounds, 
the main effects of the deformation are found in the BZ region around $\vec{k_0}=(\pi,\pi)$ of the large BZ  i.e. with one Fe/cell.\cite{raghu}, 
 as Figure \ref{deformation}(a) shows. We include results for two values of the orthorhombicity parameter $\delta = (a-b)/(a+b)$\cite{shimojima2014}, 
 namely,   $\delta=0$ and $\delta=0.002$ .  
  Our results  indicate that the energetically non-equivalent $xz$ and $yz$ orbitals\cite{raghu} become degenerate at and above the structural transition,
   in agreement with recent ARPES experiments.\cite{shimojima2014} The symmetry breaking, manifested in the band splitting appearing at $\vec{k_0}$, 
   results from the lattice deformation  from tetragonal to orthorhombic. 
Next,  Fig. \ref{deformation}(b) exhibits  the temperature dependence  we calculated for the band splitting at $\vec{k_0}$, 
measured by: $\Phi(T)=E_{d}(\vec{k_0})-E_{c}(\vec{k_0})$. Notice that $\vec{k_0}$ of the large BZ, corresponds 
to the centre of the small BZ  obtained  with two Fe/cell, i.e. $\Gamma$.
For comparison, in Fig. \ref{deformation}(b) we also include ARPES results for $\Phi(T)$  at $\Gamma$ and $M$ ( using the small BZ notation, as in ARPES\cite{shimojima2014,nakayama,watson2015}).
Ref.\cite{shimojima2014} mentions that the band splitting measured at $M$  is nearly comparable to that at $\Gamma$, 
possibly due to the relatively large error bars for these data.
The inset of Fig. \ref{deformation}(b) depicts the T-dependence of $\delta$, resulting from the T-dependent FeSe lattice parameters of Ref.\cite{khasanov2010}. 

\section{Conclusions}\label{conclusions}
We studied magnetotransport in the normal state of Fe$_{x}$Se compounds, presenting  experimental data obtained in single crystals
 as well as a theoretical description of the results. Using a simplified microscopic model to describe the compounds, based on two correlated effective orbitals, 
we determined the normal state electrical conductivity tensor and Hall coefficient in the linear response regime, employing the Kubo formulation. 
We decoupled the equations of motion for the current-current correlation functions in first-order (Hartree-Fock) approximation,  
with model parameters  in the range relevant for Fe-chalcogenides, previously used to describe their spectral properties.
With this simplified model we could successfully describe:
i) the effect of the structural transition from a tetragonal to an orthorhombic phase  observed in the ab-plane electrical resistivity; 
ii) the  positive magnetoresistance in presence of a magnetic field perpendicular to the ab-plane in the orthorhombic phase, which becomes negligible above the structural transition temperature;
iii) the Hall coefficient $R_{H}$ as a function of temperature, showing that the inclusion of moderate electron correlations  improves the description of the experimental results;
iv)  effects of the lattice deformation related to the structural transition  on the  electronic properties of FeSe superconductors: we found changes in the electronic structure below the structural phase transition temperature,  comparable to those reported in ARPES experiments.

Our work presents experimental and theoretical evidence confirming the key role of the structural transition on the strongly anisotropic magnetotransport 
properties observed in the normal state of $\beta$-FeSe superconductors, and that moderately correlated multiband models can provide the best description 
of these experimental results.

\section{Acknowledgments}

G.N. and C.I.V. are  researchers of CONICET (Argentina). M.L.A. and J.D.Q.F. have CONICET fellowships. 
We acknowledge support from CONICET (PIP 0448 and PIP 0702), ANPCyT (PICT Raices'2012, nro.1069) and SeCTyP-UNCuyo.

\end{document}